\documentclass[aps,prb,floatfix,twocolumn,superscriptaddress,showpacs,amsmath,amssymb]{revtex4-1}
\usepackage{graphicx}
\usepackage{epstopdf}
\usepackage{epsfig} 
\allowdisplaybreaks

\newcommand{\be}{\begin{equation}}
\newcommand{\ee}{\end{equation}}
\newcommand{\bea}{\begin{eqnarray}}
\newcommand{\eea}{\end{eqnarray}}
\newcommand{\ba}{\begin{eqnarray*}}
\newcommand{\ea}{\end{eqnarray*}}

\newcommand{\bw}{\begin{widetext}}
\newcommand{\ew}{\end{widetext}}

\begin{document}

\title{ Two-Band $s_\pm$  Strongly Correlated Superconductivity in K$_3$ p-Terphenyl? }

\author{Michele Fabrizio} 
\affiliation{International School for
  Advanced Studies (SISSA), Via Bonomea
  265, I-34136 Trieste, Italy} 
\author{Tao Qin}
\affiliation{Institut f\"ur Theoretische Physik, Goethe-Universit\"at, 60438 Frankfurt am Main, Germany}
\author{S. Shahab Naghavi}
\affiliation{Department of  Materials Science and Engineering,  Northwestern University, Evanston, Illinois 60208, USA}
\author{Erio Tosatti} 
\affiliation{International School for
  Advanced Studies (SISSA), Via Bonomea 265, I-34136 Trieste, Italy}
 \affiliation{CNR-IOM Democritos National Simulation Center, Via Bonomea 265, I-34136 Trieste, Italy} 
\affiliation{International  Centre   for  Theoretical  Physics (ICTP), Strada Costiera 11, I-34151 Trieste, Italy}

\date{\today} 


\begin{abstract}
A new organic superconductor, possibly with formula $K_3$ p-terphenyl, has been discovered by Wang, Gao, Huang and Chen, reaching a very high $T_c$ of about 120~K.  Besides a clear diamagnetic signal, most other details such as stoichiometry and structure are yet unknown. However, pristine p-terphenyl has a familiar  $P2_1/a$  staggered bimolecular structure, and it can be reasonably assumed that a similar bimolecular structure could be retained by hypothetical $K_3$ p-terphenyl. We point out that the resulting 2-narrow band metal would support the same $s_{\pm}$ superconductivity recently proposed for doped polycyclic aromatic hydrocarbons such as La-phenanthrene or $K_3$-picene. In that model, narrow bands, a large Hubbard $U$ and the neighbourhood of a Mott transition enhance superconductivity rather than damaging it. 

\end{abstract}
\maketitle

The interest in organic superconductors is periodically revived by the discovery of new compounds that join this wide class of materials, ranging from the earliest Bechgaard 
salts~\cite{Jerome1980}, for a review see e.g. Refs.~\onlinecite{Saito&Yoshida2011} and \onlinecite{Lang2008}, to the latest doped polycyclic aromatic hydrocarbons 
(PAHs)\cite{PAH-SC1,PAH-SC2,PAH-SC3,PAH-SC4,wang2011}. The freshest advance is the report of  Meissner diamagnetism below a temperature as high as $T_c \sim 120~\text{K}$ in potassium doped \textit{p}-terphenyl~\cite{120K}. When confirmed~\cite{Wang1,Wang2,Dessau}, this discovery will promote organic superconductors to the same rank as high $T_c$ copper oxides. 
As in other cases like e.g. over-expanded Cs$_3$C$_{60}$~\cite{Ganin,Alloul2013} fullerides, the primary question is how an organic conductor can at all superconduct, with such  enormous $T_c$, and a presumable $s$-wave pairing,  despite the necessarily large Coulomb pseudopotential $\mu_*$ compared to their narrow molecular bandwidth $W$. Na{\"i}vely this would require a pairing strength exceeding $\mu_*$, and thus also $W$. Despite suggestions of bipolaron superconductivity~\cite{120K} such a strong coupling regime is ordinarily impossible to reconcile with a $T_c$ as high as $120~\text{K}$. Besides the recent broad discussion of Baskaran\cite{Baskaran2017}, no alternative suggestions have so far appeared addressing the puzzle represented by this discovery. 

Here we discuss specifically the possibility that superconductive pairing could in this system  be strengthened rather than weakened by Coulomb repulsion. An example of that was proposed~\cite{Science2002, nostroRMP} and documented by experiments~\cite{Ganin,Alloul2013} in doped fullerides. The crux is a pairing mechanism that operates in a channel orthogonal to charge; in other words a pairing that is not a charge attraction.  In fullerides that mechanism is provided by the Jahn-Teller effect, which essentially realises the intra-molecular analogue of the resonating valence-bond (RVB) physics proposed for  high $T_c$ copper oxides.~\cite{Anderson-RVB} The C$_{60}$ orbital degeneracy allows the conduction electrons within a molecule to arrange in a low-spin dynamical Jahn Teller configuration that contains intra-molecular pairing already in the Mott insulating state; and the superconductor arises directly from metallisation of that low-spin Mott insulator.   

\begin{figure}[htb]
\begin{center}
\includegraphics[width=0.5\textwidth]{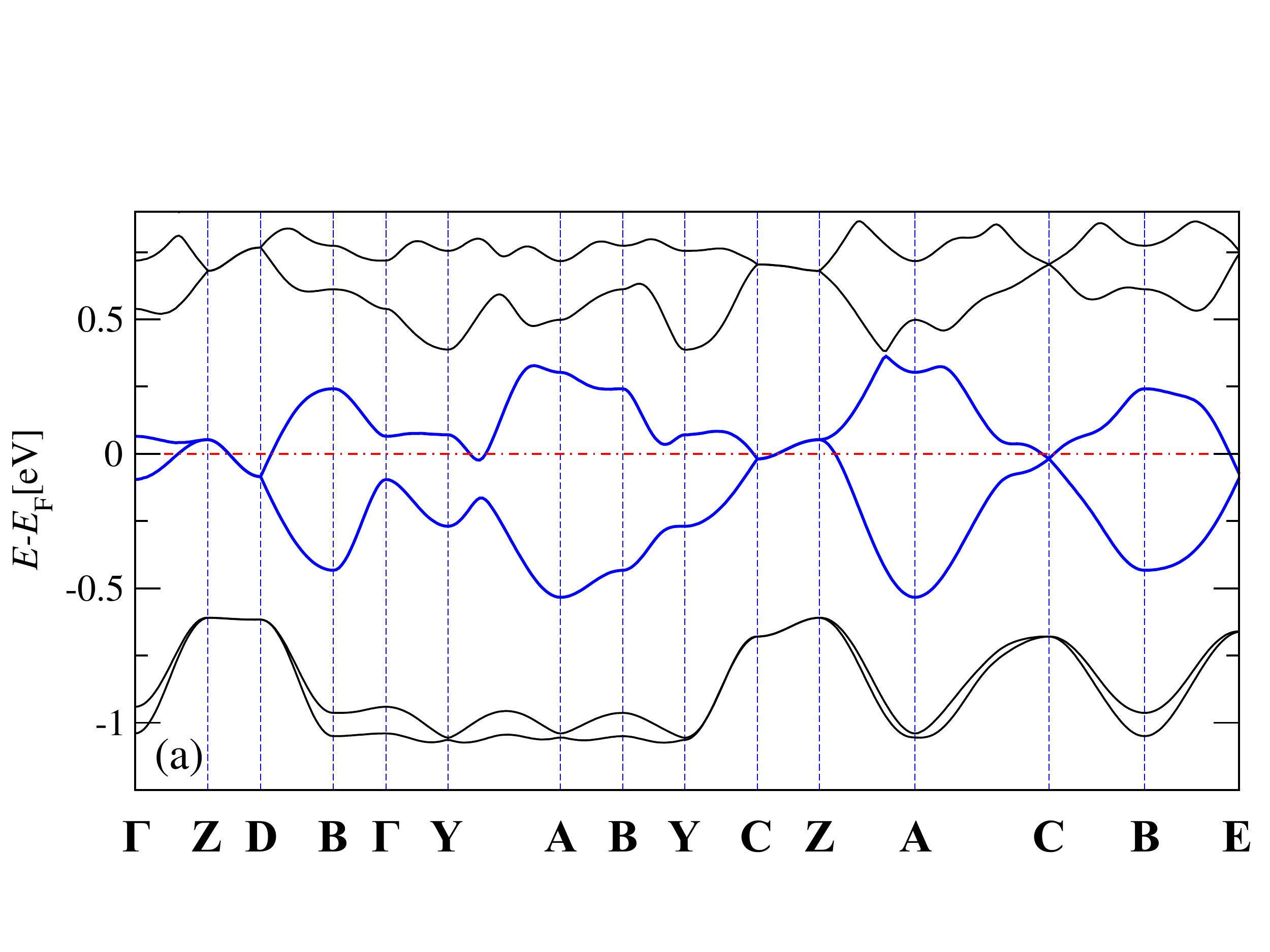}\\
\includegraphics[width=0.2\textwidth]{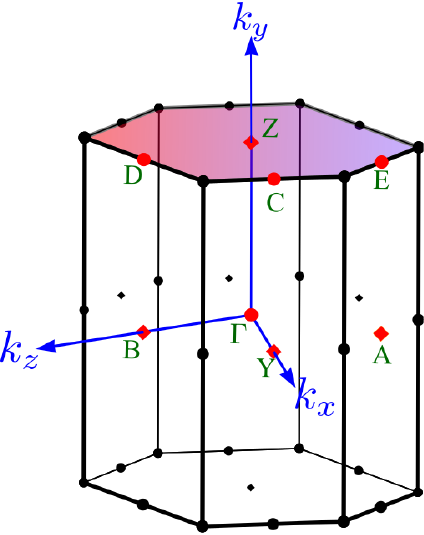}
\end{center}
\caption{Band structure (top panel) of La-doped Phenanthrene along the symmetry directions shown in the Brillouin zone (bottom panel). The zero of energy corresponds to the Fermi energy. Only the LUMO (two lowest bands), LUMO+1 (the two intermediate bands coloured in blue) and LUMO+2 (the two highest bands) are shown. On the whole upper surface of the Brillouin zone, which is coloured, each pair of bands are degenerate. }
\label{fig-1}
\end{figure}

A different but related scenario was proposed for electron-doped polycyclic aromatic hydrocarbons (PAHs) such as La-doped Phenanthrene~\cite{ShahabPRB2013,ShahabPRB2014}. A very general structural theme of these doped organics is, coarsely speaking,  a herringbone-like arrangement where two mutually staggered molecules, and not one, constitute the basic unit.  Owing to this bimolecular structure, the degeneracy that provides the required "orbital" freedom is inter-molecular rather that intra-molecular. Epitomising this element, we used a $P2_1$ lattice structure with a screw axis symmetry element, making the two molecules exactly equivalent through a mixed rotation-fractional translation symmetry operation. Four of the six electrons donated by the dopants fill completely two LUMO-derived molecular bands, and the remaining two electrons are accommodated in the two LUMO+1  bands higher in energy. Symmetry requires these two bands to be degenerate across the Fermi energy on a whole manifold at the Brillouin zone boundary, so that, despite an even number of conduction electrons per unit cell, the band structure is that of a half-filled two-band metal, as shown in Fig.~\ref{fig-1} (calculated for La-phenanthrene).  At the same time the intra-molecular Coulomb repulsion $U$ is expected to be strong enough to push the system close to a Mott insulating state. In the metallic state, such a system is highly susceptible to a inter-molecular perturbations lifting the partial degeneracy of the conduction bands, just like doped C$_{60}$ is highly responsive to the Jahn-Teller effect that splits the threefold LUMO degeneracy.  

One can envisage two different ways of lifting the degeneracy without fighting the Coulomb repulsion: 
\textit{i}) Antiferromagnetic ordering may set in, making the two molecules in the unit cell magnetically inequivalent; or \textit{ii}) a Peierls distortion may set in, prompted by the condensation of zone-boundary inter-molecular phonons, and leading to a bond-density-wave band insulating state. Either of the two symmetry breaking order parameters will also characterise a hypothetical Mott insulating phase. In the strongly correlated metal preceding, and adjacent to, that Mott state, one should therefore expect the gradual formation of local moments precursor of magnetism, and at the same time an effective enhancement of the coupling to intermolecular phonons acting to "dimerise"  the two molecules. This specific phonon was shown to mediate pairing in a $s_{\pm}$ channel, with a spin-singlet intra-band order parameter that changes sign from one band to the other. The pair essentially corresponds to singlets shared between neighbouring molecules. Since superconductivity here does not require any particular property such as nesting of the Fermi surface, but just a finite density of states at the Fermi level, we showed by a Gutzwiller calculation that an $s_{\pm}$ superconducting dome appears right before the Mott transition~\cite{ShahabPRB2014}.  Strikingly, that superconducting state is not hampered, but is on the contrary supported, by the intramolecular Coulomb repulsion $U$. The physics of that two-band Hubbard-Fr\"ohlich model~\cite{ShahabPRB2014}, with the due differences, is of the same family to that proposed for superconducting fullerides~\cite{nostroRMP}, and, ultimately, analogue to the RVB scenario for high $T_c$ copper oxides~\cite{Anderson-RVB}.   \\

Pristine \textit{p}-terphenyl crystallises in a structure with $P2_1/a$ symmetry~\cite{Rietveld1970}, with two staggered molecules per cell. Ab initio structural searches have reasonably suggested that even in doped PAHs that structural motif is retained. If indeed the superconducting compound of Wang \textit{et al.}~\cite{120K} was confirmed to be K$_3$ \textit{p}-terphenyl, and if it preserved a staggered bimolecular structure, all what was said about La-Phenanthrene model would apply here too. Indeed recent electronic structure calculations~\cite{Gao-bande} show the same degeneracy of the two LUMO+1 bands at the Fermi level that we demonstrated in  La-Phenanthrene~\cite{ShahabPRB2013}. In the light of these similarities, we propose that the superconducting order parameter in  K$_3$ \textit{p}-terphenyl might also possess $s_\pm$ symmetry with a pairing mechanism mediated by the coupling to dimerising phonons enhanced by the proximity to a Mott transition. Our estimate of $T_c$ in La-Phenanthrene was however quite low, comparable to the earliest evidence of a Meissner effect in K$_3$ \textit{p}-terphenyl below 7~K~\cite{Wang1}, rather than the later ones of 43~K~\cite{Wang2} and 120~K~\cite{120K}. More recent photoemission data show a gap opening at 63~K, although that does not necessarily imply onset of superconductivity. Therefore more extensive theoretical and experimental work is required to firmly confirm superconductivity at such high temperatures and clarify its origin.  We are grateful to G. Baskaran for pointing out the existence of this system.

\bibliographystyle{apsrev}
\bibliography{twoband.bib}

\end{document}